\documentclass[conference,a4paper]{APSIPA2026}
\usepackage{amsmath}
\usepackage{graphicx}
\usepackage{multirow}
\usepackage{booktabs}
\usepackage{makecell}
\usepackage{array}
\usepackage{amsmath,amssymb,bm}
\usepackage{bbm}
\usepackage{threeparttable}
\usepackage{listings}
\usepackage{latexsym}
\usepackage{subcaption}
\usepackage{algorithm}
\usepackage{makecell}
\usepackage{xcolor}
\usepackage{algpseudocode}
\usepackage{siunitx}
\usepackage[backend=biber,style=ieee,]{biblatex}
\usepackage[capitalise,noabbrev]{cleveref}

\crefname{figure}{Fig.}{Figs.}
\Crefname{figure}{Fig.}{Figs.}

\crefname{table}{Table}{Tables}
\Crefname{table}{Table}{Tables}

\crefname{section}{Section}{Sections}
\Crefname{section}{Section}{Sections}
\crefname{subsection}{Section}{Sections}
\Crefname{subsection}{Section}{Sections}
\crefname{subsubsection}{Section}{Sections}
\Crefname{subsubsection}{Section}{Sections}

\crefname{appendix}{Appendix}{Appendices}
\Crefname{appendix}{Appendix}{Appendices}

\crefname{equation}{}{}
\Crefname{equation}{Equation}{Equations}

\crefformat{equation}{(#2#1#3)}
\Crefformat{equation}{Equation~(#2#1#3)}

\crefrangeformat{equation}{(#3#1#4)--(#5#2#6)}
\Crefrangeformat{equation}{Equations~(#3#1#4)--(#5#2#6)}

\crefmultiformat{equation}{(#2#1#3)}%
  { and~(#2#1#3)}%
  {, (#2#1#3)}%
  {, and~(#2#1#3)}

\Crefmultiformat{equation}{Equations~(#2#1#3)}%
  { and~(#2#1#3)}%
  {, (#2#1#3)}%
  {, and~(#2#1#3)}

\usepackage{geometry}
\usepackage{fancyhdr}

\fancypagestyle{firststyle}{
  \fancyhf{}
  \fancyhead[C]{2026 Asia Pacific Signal and Information Processing Association Annual Summit and Conference (APSIPA ASC)}
}
\usepackage{xparse}
\DeclareSIUnit{\cent}{cent}
\NewDocumentCommand\newletter{mmomm}{%
\NewDocumentCommand#1{st@o}{%
\IfBooleanTF{##1}{\mathbf{\MakeUppercase{#2}}\IfValueT{#3}{^{#3}}}{%
\IfBooleanTF{##2}{\mathbf{#2}\IfValueT{#3}{^{#3}}_{\IfValueTF{##3}{##3}{#5}}}{%
{#2}\IfValueT{#3}{^{#3}}_{\IfValueTF{##3}{##3}{#4}}%
}}}}

\NewDocumentCommand\newletterbm{mmomm}{%
\NewDocumentCommand#1{st@o}{%
\IfBooleanTF{##1}{\bm{\MakeUppercase{#2}}\IfValueT{#3}{^{#3}}}{%
\IfBooleanTF{##2}{\bm{#2}\IfValueT{#3}{^{#3}}_{\IfValueTF{##3}{##3}{#5}}}{%
{#2}\IfValueT{#3}{^{#3}}_{\IfValueTF{##3}{##3}{#4}}%
}}}}

\NewDocumentCommand{\newvec}{m m m o}{%
  \NewDocumentCommand#1{s t! o}{%
    \IfBooleanTF{##1}
      {%
        \bm{#3}%
        \IfValueT{##3}{_{##3}}%
      }%
      {%
        \bm{#2}%
        \IfBooleanF{##2}{%
          \IfValueTF{##3}
            {_{##3}}
            {\IfValueT{#4}{_{#4}}}%
        }%
      }%
  }%
}

\NewDocumentCommand{\newdecorvec}{m m m m o}{%
  \NewDocumentCommand#1{s t! o}{%
    \IfBooleanTF{##1}
      {%
        #4{\bm{#3}}%
        \IfValueT{##3}{_{##3}}%
      }%
      {%
        #4{\bm{#2}}%
        \IfBooleanF{##2}{%
          \IfValueTF{##3}
            {_{##3}}
            {\IfValueT{#5}{_{#5}}}%
        }%
      }%
  }%
}

\def\R{\mathbb{R}}
\def\Rnn{\mathbb{R}_{\geq0}}

\def\msloss{\mathcal{L}_{\text{MS}}}
\def\Dec{\text{Dec}}
\def\fscore{f^{\text{(score)}}}
\def\lscore{l^{\text{(score)}}}
\def\zscore{z^{\text{(score)}}}
\def\Zscore{Z^{\text{(score)}}}
\def\diffsteps{T^\mathrm{(DDPM)}}
\def\reg{\mathcal{R}}

\def\Lprop{\mathcal{L}_{\text{prop}}}

\newvec{\vs}{s}{s}[k]
\newvec{\vx}{x}{x}[t]
\newvec{\vz}{z}{z}[t]
\newvec{\vf}{f}{f}[t]
\newvec{\vl}{l}{l}[t]
\newdecorvec{\vshat}{s}{s}{\hat}[k]
\newvec{\vy}{y}{y}
\newdecorvec{\vyhat}{y}{y}{\hat}
\newvec{\vp}{p}{p}[k,t]
\newvec{\vfscore}{\fscore}{\fscore}[k,t]
\newvec{\vlscore}{\lscore}{\lscore}[k,t]
\newvec{\vzscore}{\zscore}{\Zscore}[k,t,d]
\newvec{\vu}{u}{u}[\tau]
\newvec{\vv}{v}{v}[\tau]
\newdecorvec{\vuhat}{u}{u}{\hat}[\tau]
\newdecorvec{\vxhat}{x}{x}{\hat}[\tau]
\newdecorvec{\vutilde}{u}{u}{\tilde}[k,\tau]
\newdecorvec{\vxtilde}{x}{x}{\tilde}[k,\tau]

\def\Underline{\setbox0\hbox\bgroup\let\\\endUnderline}
\def\endUnderline{\vphantom{y}\egroup\smash{\underline{\box0}}\\}
\def\|{\verb|}
\def\Fo{F_{\text{o}}}

\begin{document}

\title{Differentiable Digital Signal Processing Mixture Model-Guided Diffusion for Synthesis Parameter Estimation from Harmonic Sound Mixtures}

\author{
\authorblockN{
Kengo Takemoto\authorrefmark{1}\authorrefmark{2} and
Tomohiko Nakamura\authorrefmark{2} and
Hiroshi Saruwatari\authorrefmark{1}
}

\authorblockA{
\authorrefmark{1}
The University of Tokyo, Tokyo, Japan
}

\authorblockA{
\authorrefmark{2}
National Institute of Advanced Industrial Science and Technology (AIST), Tokyo, Japan
}
}

\maketitle
\begingroup
\renewcommand\thefootnote{}
\footnotetext{This work was supported by JSPS KAKENHI Grant Number JP23K28108.}
\endgroup
\thispagestyle{firststyle}
\pagestyle{empty}

\begin{abstract}
A differentiable digital signal processing (DDSP) autoencoder reconstructs a monophonic harmonic sound through three types of synthesis parameters: fundamental frequency, loudness, and timbre features.
To handle mixtures of harmonic sounds within the DDSP approach, we have previously proposed a DDSP mixture model (DDSPMM).
It represents a mixture as the sum of source signals synthesized by the decoders of pretrained DDSP autoencoders.
Although DDSPMM enables direct estimation of synthesis parameters of each source from mixtures,
it does not explicitly model temporal variations in the synthesis parameters and can produce excessive temporal fluctuations.
In this paper, we propose a method for estimating synthesis parameters with temporally plausible trajectories by incorporating a denoising diffusion probabilistic model (DDPM) into the DDSPMM-based estimation.
The DDPM is trained as a generative model of synthesis parameters.
During estimation, the proposed method guides the DDPM reverse diffusion process with the reconstruction error between the observed mixture and the mixture synthesized by DDSPMM from the current estimates.
Experiments on woodwind and string instrument ensembles showed that the DDPM-based regularization improves synthesis parameter estimation by imposing temporal plausibility on the estimated trajectories.
\end{abstract}

\section{Introduction}
Neural audio synthesis has achieved high-quality generation of musical instrument sounds (e.g.,~\cite{Engel2017ICML,Rouard2021CRASH,Ben2024FSP}).
Beyond sound quality, applications such as expressive performance rendering~\cite{Wu2022MIDIDDSP,Kim2024ICASSP} and timbre transfer~\cite{Caillon2021arXiv,Mancusi2025ICASSP} require synthesis models with controllable and interpretable representations.
However, purely data-driven synthesis models often rely on latent representations that are difficult to interpret or manipulate in terms of musically meaningful attributes.

One promising approach for controllable and interpretable musical instrument sound synthesis is differentiable digital signal processing (DDSP)~\cite{Ben2024FSP}.
It implements signal processing modules in a form amenable to backpropagation and combines them with neural networks (NNs).
A representative model for monophonic harmonic instrument sounds is the DDSP autoencoder~\cite{Engel2020ICRL}.
It represents an input signal using frame-wise parameters: fundamental frequency ($\Fo$), loudness, and timbre features.
We refer to them as synthesis parameters in this paper.
The decoder then reconstructs the signal from these parameters using additive and subtractive synthesizers.

To handle a mixture of multiple harmonic instrument sounds, we have previously proposed a DDSP mixture model (DDSPMM)~\cite{Kawamura2022ICASSP}.
It represents a mixture as the sum of source signals synthesized by the decoders of pretrained DDSP autoencoders.
This representation enables direct estimation of the synthesis parameters of each source from an observed mixture, without separating the mixture into individual sources.
In this procedure, the synthesis parameters are initialized using musical score information and refined by fitting the synthesized mixture to the observed mixture.

Despite this capability, the DDSPMM-based estimation method tends to produce synthesis parameters with excessive temporal fluctuations, as we will show in \Cref{sec:results}.
During fitting, the synthesis parameters are treated as independent variables at each time frame, and temporal constraints on synthesis parameter trajectories are not explicitly imposed.
However, synthesis parameters observed in real performances do not vary arbitrarily over time.
For example, in sustained regions of instrument sounds, $\Fo$ and timbre features tend to vary smoothly, whereas loudness and timbre features can change more quickly around note onsets.
Ignoring these temporal characteristics can result in trajectories that deviate from those observed in real performances.

In this paper, we propose a DDSPMM-based method for estimating synthesis parameters with temporally plausible trajectories.
Specifically, we incorporate a generative model of synthesis parameter trajectories into the DDSPMM-based fitting procedure.
As the generative model, we use a denoising diffusion probabilistic model (DDPM)~\cite{ho2020denoising}, pretrained to generate synthesis parameter trajectories conditioned on musical score information.
During the reverse diffusion process of DDPM, the proposed method uses DDSPMM to resynthesize the mixture from the current synthesis parameter estimates and guides these estimates using the reconstruction error between the synthesized and observed mixtures.
Experiments show that the proposed method improves estimation accuracy, especially for loudness and timbre features.

\section{Related work}
\subsection{DDSP Autoencoder}
The DDSP autoencoder consists of an encoder and a decoder~\cite{Engel2020ICRL}.
Given an input signal $\vs! \in \R^N$ of length $N$, the encoder extracts synthesis parameters $\vx*\in\R^{T\times(D+2)}$ over $T$ time frames.
The $t$-th row of $\vx*$ is given by $[f_t,l_t,\vz^\top]$, where $t=1,\ldots,T$ is the time frame index.
The $\Fo$, denoted by $f_t\in\Rnn$, is estimated using the NN-based pitch estimator CREPE~\cite{kim2018crepe}.
The loudness $l_t\in\R$ is computed by averaging the power spectrum over frequency bins and taking its logarithm.
The timbre feature vector $\vz\in\R^D$ is computed from mel-frequency cepstral coefficients (MFCCs) using a gated recurrent NN.
The decoder generates a synthesized signal $\vshat!\in\R^{N}$ from $\vx*$ using additive and subtractive synthesizers and an optional reverb module.
It estimates the control parameters for the synthesizers, while the reverb module is implemented as a convolutional layer.
The differentiability of these modules allows gradients to be propagated through the entire network during training.

The DDSP autoencoder is trained by minimizing the multi-scale spectral loss~\cite{Engel2020ICRL} between $\vs!$ and $\vshat!$.
For $M$ time-frequency resolutions, let $m=1,\ldots,M$ be the time-frequency resolution index and $\mathcal{S}_m$ denote the magnitude short-time Fourier transform (STFT) at resolution $m$.
The loss is defined as
\begin{equation}
\begin{aligned}
\msloss(\vs!,\vshat!)
&:=
\sum_{m=1}^M
\Bigl(
\left\lVert
\mathcal{S}_m(\vs!)
-
\mathcal{S}_m(\vshat!)
\right\rVert_1\\
&\qquad\qquad
+
\left\lVert
\log(\mathcal{S}_m(\vs!))
-
\log(\mathcal{S}_m(\vshat!))
\right\rVert_1
\Bigr),
\end{aligned}
\end{equation}
where $\lVert\cdot\rVert_1$ denotes the $\ell_1$ norm and $\log$ is the element-wise logarithm.

\subsection{Synthesis Parameter Estimation Using DDSPMM}
\label{sec:ddspmm}
DDSPMM represents the generation process of a mixture of $K$ monophonic instrument sounds from source-wise synthesis parameters $\vx*[k]\in\R^{T\times(D+2)}$~\cite{Kawamura2022ICASSP}, where $k=1,\ldots,K$ indexes the sources.
Each source signal is generated by the decoder of a pretrained DDSP autoencoder, referred to as the source synthesizer.
The synthesized mixture is given by the sum of these source signals.

Given an observed mixture $\vy\in\R^{N}$, estimating $\{\vx*[k]\}_k$ is formulated as the inverse problem of this generation process:
\begin{equation}
\label{eq:mixture-ddsp-opt}
\min_{\left\{\vx*[k]\right\}_{k=1}^{K}}
\msloss \left( \vy, \sum_{k=1}^K\Dec_{\psi_k}(\vx*[k]) \right),
\end{equation}
where $\Dec_{\psi_k}$ denotes the pretrained source synthesizer with fixed parameters $\psi_k$. 
Owing to the differentiability of $\Dec_{\psi_k}$, $\{\vx*[k]\}_k$ can be iteratively updated by gradient descent.

When musical score information is available, it can provide pitch and note-activity cues for initializing the synthesis parameters.
Assume that the score information is given in Musical Instrument Digital Interface (MIDI) format.
The initial $\Fo$ value, $\fscore_{k,t}$, is defined as
\begin{equation}
\label{eq:f0_score_initialization}
\fscore_{k,t}=
\begin{cases}
    440\times2^{(p_{k,t}-69)/12} & (p_{k,t}\geq0 ) \\
    440\times2^{(\bar{p}_k-69)/12} & (p_{k,t}=-1)
\end{cases},
\end{equation}
where $p_{k,t}\in\{-1,0,\ldots,127\}$ is the MIDI note number for source $k$ at frame $t$, $p_{k,t}=-1$ indicates silence, and $\bar{p}_k$ is the average MIDI note number over note-active regions of source $k$ in the musical score.
The initial loudness value, $\lscore_{k,t}$, is defined as
\begin{equation}
\lscore_{k,t} =
\begin{cases}
    l_\mathrm{high} & (p_{k,t}\geq0) \\
    l_\mathrm{low} & (p_{k,t}=-1)
\end{cases},
\label{eq:ld_score_initialization}
\end{equation}
where $l_\mathrm{high}$ and $l_\mathrm{low}$ are predetermined values for frames with active and inactive notes, respectively.
With this score-informed initialization, the DDSPMM-based analysis-by-synthesis method has been shown to improve estimation accuracy compared with a separation-and-analysis method that applies the DDSP autoencoder after source separation~\cite{Kawamura2022ICASSP}.

\subsection{DDPM}
\label{sec:ddpm}
DDPM consists of a fixed diffusion process that gradually maps data to standard normal noise and a learned reverse diffusion process~\cite{ho2020denoising}.
Our method uses conditional DDPM, in which the reverse diffusion process is conditioned on auxiliary information $\bm{c}$.
We briefly review the formulation used in this paper; see~\cite{ho2020denoising} for a detailed derivation.

In the diffusion process, standard normal noise is gradually added to a clean variable $\vu[0]$ over $\diffsteps$ steps.
Let $\tau=1,\ldots,\diffsteps$ denote the diffusion step and $\beta_{\tau}\in(0,1)$ be the noise schedule.
Defining $\alpha_{\tau}=1-\beta_{\tau}$ and $\bar{\alpha}_{\tau}=\prod_{\tau^\prime=1}^{\tau}\alpha_{\tau^\prime}$, the noisy variable at step $\tau$ can be sampled directly from $\vu[0]$:
\begin{equation}
\label{eq:diffusion-process}
\vu[\tau]
=
\sqrt{\bar{\alpha}_{\tau}}\vu[0]
+
\sqrt{1-\bar{\alpha}_{\tau}}\,\bm{\varepsilon},
\end{equation}
where each element of $\bm{\varepsilon}$ is independently sampled from the standard normal distribution.

To learn the reverse diffusion process, DDPM uses a noise predictor $\bm{\varepsilon}_{\theta}$ parameterized by $\theta$.
It is trained to predict the added noise from $\vu[\tau]$, $\tau$, and $\bm{c}$ by minimizing
\begin{equation}
\mathcal{L}_{\mathrm{DDPM}}(\theta)
=
\mathbb{E}_{(\vu[0],\bm{c}),\bm{\varepsilon},\tau}
\left[
\left\lVert
\bm{\varepsilon}
-
\bm{\varepsilon}_{\theta}(\vu[\tau],\tau,\bm{c})
\right\rVert_2^2
\right].
\label{eq:DDPM_loss}
\end{equation}
Here, $(\vu[0],\bm{c})$ is sampled from training data and $\tau$ is uniformly sampled from $\{1,\ldots,\diffsteps\}$.

The noise-prediction objective of DDPM can also be interpreted as a form of score function estimation~\cite{ho2020denoising}.
For the conditional distribution of noisy data $p_{\tau}(\vu[\tau]\mid\bm{c})$ at diffusion step $\tau$, the score function is given by
$\nabla_{\vu[\tau]}\log p_{\tau}(\vu[\tau]\mid\bm{c})$.
In the noise-prediction formulation, the trained network provides the following score estimate:
\begin{equation}
\nabla_{\vu[\tau]}\log p_{\tau}(\vu[\tau]\mid\bm{c})
\approx
-\frac{\bm{\varepsilon}_{\theta}(\vu[\tau],\tau,\bm{c})}{\sqrt{1-\bar{\alpha}_{\tau}}}.
\end{equation}
Since the reverse diffusion process is driven by this score estimate, a gradient of an external objective can be added to guide the generated sample during denoising.

\section{Proposed Method} \label{sec:proposed}
The proposed method incorporates a generative model of synthesis parameter trajectories into DDSPMM-based estimation.
We assume the same setting as in \cite{Kawamura2022ICASSP}, where 
time-aligned musical score information is available for the observed mixtures.
This section describes the generative model and its integration with the estimation procedure.

\begin{figure}[t]
\centering
\includegraphics[width=\linewidth]{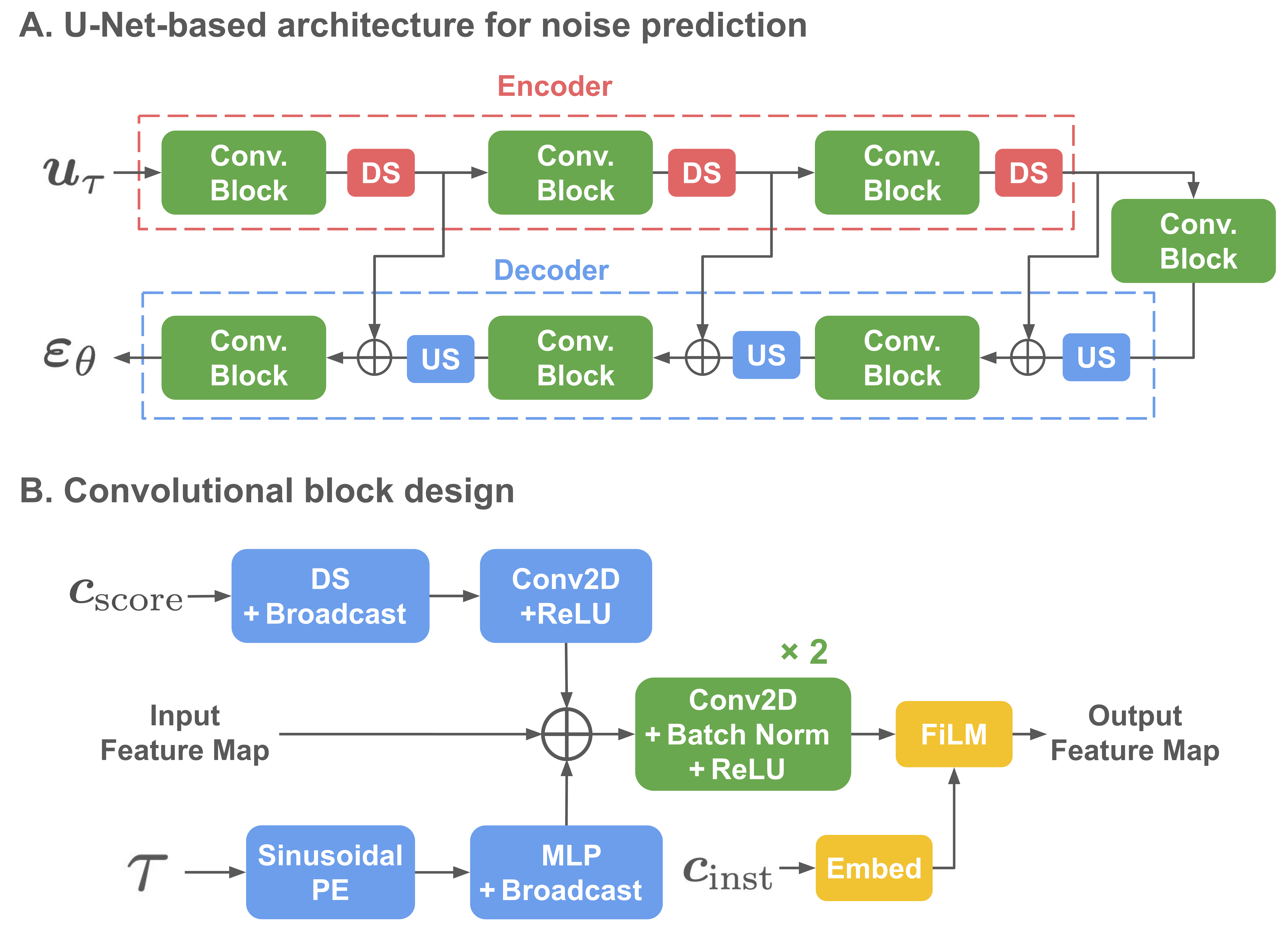}
\caption{
(A): noise predicter based on the U-Net architecure. 
DS and US denotes downsampling and upsampling, respectively.
(B): convolutional block used in the noise predictor.
PE and Conv2D denote positional encoding and two-dimensional convolutional layer, respectively.
}
\label{fig:unet}
\end{figure}

\begin{figure*}[t]
\centering
\includegraphics[width=0.9\linewidth]{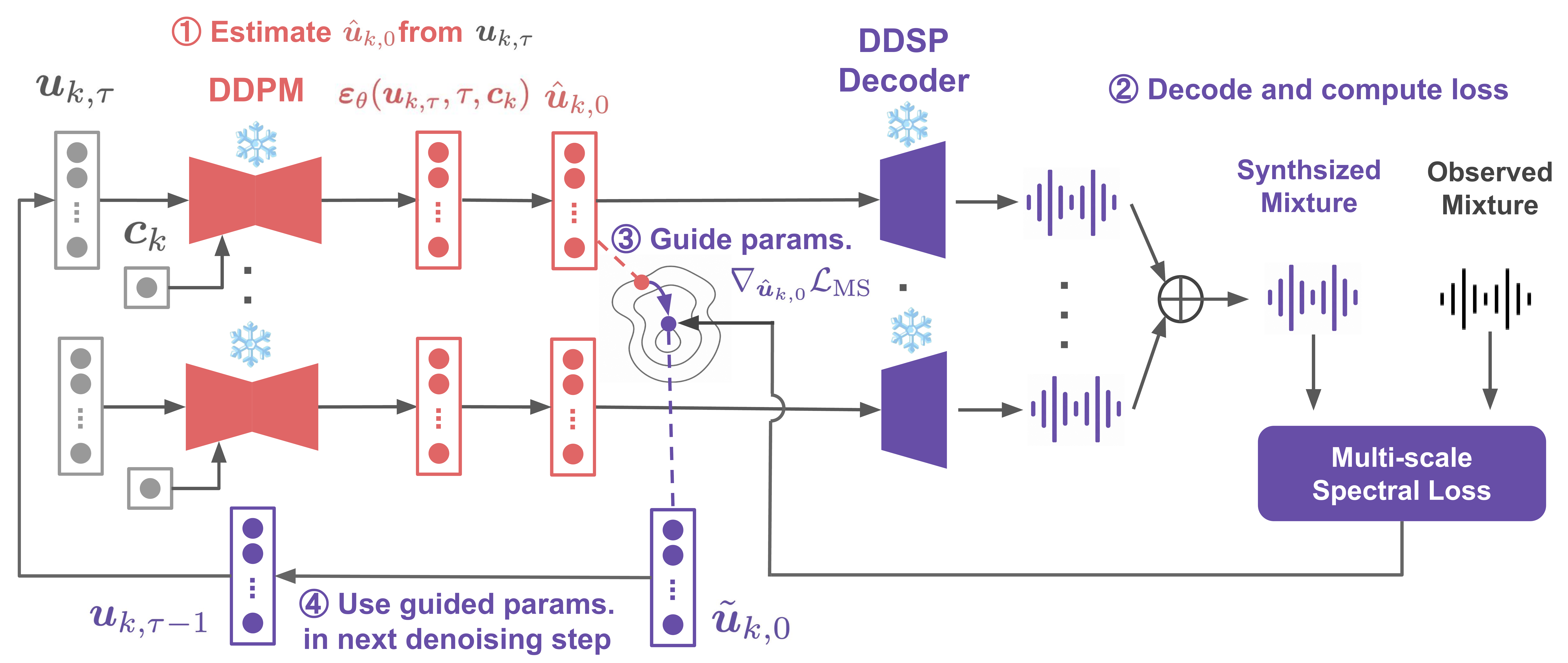}
\caption{Overview of the proposed synthesis parameter estimation algorithm.
At each step, reverse diffusion is guided by the multi-scale spectral loss between the observed mixture and the mixture synthesized from the current estimates.}
\label{fig:ddpm-guidance}
\end{figure*}

\subsection{DDPM-Based Generative Model of Synthesis Parameters}
\label{sec:ddpm-generation}
We use DDPM as the generative model of synthesis parameters and condition it on musical score information, including pitch, note activity, and instrument type.
This conditioning allows the DDPM to model temporal variations in synthesis parameters associated with the musical score.
In the following, we reuse the notation from \Cref{sec:ddpm}.
For brevity, we omit the source index $k$ since the DDPM is trained on isolated instrument performances.

\noindent \textbf{DDPM target:}
For $\Fo$ and loudness, the DDPM generates residuals with respect to the score-informed initial values defined in \Cref{eq:f0_score_initialization,eq:ld_score_initialization}.
In preliminary experiments, we observed that this residual representation led to more stable modeling of pitch transitions and loudness dynamics than directly generating raw $\Fo$ and loudness.
The $\Fo$ residuals, loudness residuals, and each dimension of the timbre features are standardized over all frames in the training data so that they have zero mean and unit variance.
The standardized components are then concatenated to form the clean DDPM variable $\vu[0]\in\R^{T\times(D+2)}$.
We denote the transform from the synthesis parameter matrix $\vx*$ to $\vu[0]$ by $\Phi$, i.e., $\vu[0]=\Phi(\vx*)$, and denote its inverse by $\Phi^{-1}$.

\noindent \textbf{Conditioning:}
The conditioning inputs are the one-hot vector of instrument type $\bm{c}_{\text{inst}}\in\{0,1\}^{I}$ and the standardized score-informed $\Fo$ sequence $\bm{c}_{\text{score}}\in\R^{T}$
given by \cref{eq:f0_score_initialization}.
Here, $I$ is the number of instrument classes.
The standardization is performed using the mean and standard deviation computed from the score-informed $\Fo$ sequences over all frames in the training data.
The noise predictor $\bm{\varepsilon}_{\theta}$ takes these variables as conditions and is trained with the loss defined in \cref{eq:DDPM_loss}.

\noindent \textbf{Network architecture:}
The noise predictor is based on the U-Net architecture~\cite{ronneberger2015unet}.
It follows an three-level encoder--decoder architecture with skip connections (see \Cref{fig:unet}-A).
Each encoder stage consists of a two-dimensional (2D) convolutional block followed by max pooling with a downsampling factor of 2.
The decoder mirrors the encoder, replacing max pooling with bilinear upsampling.
The convolutional block is additionally inserted at the bottleneck between the encoder and decoder.
\Cref{fig:unet}-B shows the convolutional block.
The diffusion step $\tau$ is encoded by sinusoidal positional encoding, transformed by a multilayer perceptron (MLP), and added to the input feature map.
The score condition $\bm{c}_{\text{score}}$ is downsampled to match the feature-map resolution.
It is then transformed by a 2D convolutional layer with a rectified linear unit (ReLU) and added to the feature map.
The resulting feature map is passed through two 2D convolutional layers, each followed by batch normalization and ReLU.
Finally, feature-wise linear modulation (FiLM)~\cite{perez2018film} is applied using channel-wise scaling and shift coefficients generated from $\bm{c}_{\text{inst}}$.

\subsection{Integration of DDPM with DDSPMM-Based Synthesis Parameter Estimation}
\label{sec:ddpm-ddspmm-integration}
After training the DDPM, we integrate it with DDSPMM-based synthesis parameter estimation.
To distinguish sources, we use the source index $k$ for source-wise variables and transforms, namely $\vx*[k]$, $\vu[k,0]$, $\vu[k,\tau]$, $\bm{c}_{\text{inst},k}$, $\bm{c}_{\text{score},k}$, and $\Phi_k$.
The condition $\bm{c}_k$ is obtained by concatenating $\bm{c}_{\text{inst},k}$ and $\bm{c}_{\text{score},k}$.

To derive the proposed method, we extend the DDSPMM-based estimation problem in \cref{eq:mixture-ddsp-opt} by introducing a regularizer $\reg_k$ on the synthesis parameters:
\begin{equation}
\label{eq:proposed-objective}
\Lprop(\{\vx*[k]\}_{k}) := \lambda \msloss\left(\vy,\sum_{k}\Dec_{\psi_k}(\vx*[k])\right) + \sum_{k} \reg_k(\vx*[k])
\end{equation}
where $\lambda\geq 0$ controls the weight of the multi-scale spectral loss.
The source synthesizers $\{\Dec_{\psi_k}\}_{k=1}^{K}$ are pretrained and fixed during estimation.

As the regularizer $\reg_k$, we consider the negative log-density of the synthesis parameters under the source-wise condition $\bm{c}_k$, i.e.,
$\reg_k(\vx*[k])=-\log p(\vx*[k]\mid\bm{c}_k)$.
The gradient of this regularizer with respect to $\vx*[k]$ is given by
\begin{equation}
\label{eq:diff-reg}
\nabla_{\vx*[k]}\reg_k(\vx*[k])
=
-\nabla_{\vx*[k]}
\log p(\vx*[k]\mid\bm{c}_k).
\end{equation}
This gradient corresponds to the negative score function of the conditional distribution.
However, this score function is generally intractable because the conditional distribution $p(\vx*[k]\mid\bm{c}_k)$ is not available in closed form.

We thus use the trained conditional DDPM to obtain a score-based direction.
As described in \Cref{sec:ddpm}, the noise predictor of DDPM provides a score estimate at each reverse diffusion step.
Rather than directly computing the regularization gradient on the clean synthesis parameter matrix $\vx*[k]$, we use the DDPM score estimate for the noisy variable $\vu[k,\tau]$ during reverse diffusion process.
This score estimate guides $\vu[k,\tau]$ toward temporal structures learned from the training data under the condition $\bm{c}_k$.
The reconstruction error $\msloss\left(\vy,\sum_{k}\Dec_{\psi_k}(\vx*[k])\right)$ is then used as an additional guidance term to make the synthesized mixture consistent with the observed mixture.

\subsection{DDSPMM-Guided Reverse Diffusion}
\label{sec:ddpm-guidance}
We now present the guided reverse diffusion algorithm for estimating source-wise synthesis parameters from an observed mixture.
At each diffusion step, the clean DDPM variable estimate is corrected using the multi-scale spectral loss between the observed mixture and the mixture synthesized by DDSPMM using pretrained source synthesizers.

At $\tau=\diffsteps$, each entry of $\vu[k,\diffsteps]$ is independently sampled from a standard normal distribution.
At each reverse diffusion step $\tau$, the clean DDPM variable is estimated as
\begin{equation}
\label{eq:vuhat0_est}
\vuhat[k,0]=\cfrac{\vu[k,\tau]-\sqrt{1-\bar{\alpha}_{\tau}}\,\bm{\varepsilon}_{\theta}(\vu[k,\tau],\tau,\bm{c}_k)}{\sqrt{\bar{\alpha}_{\tau}}}.
\end{equation}
Using the corresponding synthesis parameter estimates obtained by $\Phi_k^{-1}$, we correct $\vuhat[k,0]$ using the gradient of the multi-scale spectral loss:
\begin{equation}
\label{eq:guided-u0}
\vutilde[k,0]
=
\vuhat[k,0]
-
\lambda
\nabla_{\vuhat[k,0]}
\msloss
\left(
\vy,
\sum_{k^\prime=1}^{K}
\Dec_{\psi_{k^\prime}}
\left(
\Phi_{k^\prime}^{-1}\left(\vuhat[k^\prime,0]\right)
\right)
\right).
\end{equation}
Using the corrected estimate $\vutilde[k,0]$, the mean for the previous diffusion step is computed as
\begin{equation}
\label{eq:mu_est}
\tilde{\bm{\mu}}_{k,\tau-1}
=
\frac{
\sqrt{\bar{\alpha}_{\tau-1}}\beta_{\tau}
}{
1-\bar{\alpha}_{\tau}
}
\tilde{\bm{u}}_{k,0}
+
\frac{
\sqrt{\alpha_{\tau}}(1-\bar{\alpha}_{\tau-1})
}{
1-\bar{\alpha}_{\tau}
}
\bm{u}_{k,\tau}.
\end{equation}
Then, $\vu[k,\tau-1]$ is sampled as
\begin{equation}
\label{eq:u_tau-1_est}
\bm{u}_{k,\tau-1}
=
\tilde{\bm{\mu}}_{k,\tau-1}
+
\sqrt{
\frac{
(1-\alpha_{\tau})(1-\bar{\alpha}_{\tau-1})
}{
1-\bar{\alpha}_{\tau}
}
}
\,\bm{\xi}_{k,\tau-1},
\end{equation}
where each entry of $\bm{\xi}_{k,\tau-1}\in\R^{T\times(D+2)}$ is independently sampled from the standard normal distribution.
By iterating \cref{eq:vuhat0_est,eq:guided-u0,eq:mu_est,eq:u_tau-1_est} until $\tau=1$, we obtain the final estimate $\vxhat*[k]$ for each source from the resulting estimate of $\vu[k,0]$.

\begin{table*}[t!]
\centering
{
\caption{Estimation errors and Zimtohrli scores for the synthesis parameter estimation methods. The Instruments column lists the instrument types in each mixture; the labels in parentheses indicate the corresponding musical pieces in the URMP dataset.}
\begin{tabular}{cclcccc}
\toprule
Instruments
&Duration\,[\si{\second}]
&Method
&$\Fo$\,[\si{\cent}] ($\downarrow$)&Loudness\,[\si{\decibel}] ($\downarrow$)&MFCCs ($\downarrow$)&Zimtohrli score ($\uparrow$)\\
\midrule
\multirow[c]{4}{*}{\makecell{Flute / Flute\\(Fugue)}}
&\multirow[c]{4}{*}{172}
&SS--DDSP
&$281\pm165$&$6.11\pm0.330$&$3.70\pm0.457$&$2.65\pm0.115$\\
&&DDSPMM
&$18.5\pm6.39$&$5.35\pm0.368$&$2.09\pm0.0860$&$3.70\pm0.0251$\\
&&Proposed
&$\bm{14.0\pm5.97}$&$\bm{4.08\pm0.833}$&$\bm{2.01\pm0.0211}$&$\bm{3.81\pm0.0575}$\\
\cmidrule(lr){3-7}
&&Reconst.
&-&-&$1.63\pm0.0400$&$4.05\pm0.00784$\\
\midrule
\multirow[c]{4}{*}{\makecell{Clarinet / Violin\\(Pavane)}}
&\multirow[c]{4}{*}{133}
&SS--DDSP
&$12.5\pm8.95$&$9.07\pm3.64$&$3.29\pm0.101$&$3.17\pm0.180$\\
&&DDSPMM
&$\bm{9.82\pm2.18}$&$5.70\pm1.22$&$2.12\pm0.146$&$\bm{3.77\pm0.159}$\\
&&Proposed
&$12.8\pm1.39$&$\bm{3.52\pm0.517}$&$\bm{2.05\pm0.0200}$&$3.71\pm0.0758$\\
\cmidrule(lr){3-7}
&&Reconst.
&-&-&$1.89\pm0.203$&$3.74\pm0.160$\\
\midrule
\multirow[c]{4}{*}{\makecell{Flute / Oboe / Clarinet\\(Fugue)}}
&\multirow[c]{4}{*}{172}
&SS--DDSP
&$\bm{14.2\pm6.41}$&$14.9\pm1.10$&$3.47\pm0.179$&$3.21\pm0.101$\\
&&DDSPMM
&$16.5\pm4.37$&$7.58\pm0.547$&$2.20\pm0.0246$&$3.66\pm0.0125$\\
&&Proposed
&$16.1\pm2.71$&$\bm{3.87\pm0.569}$&$\bm{2.02\pm0.0764}$&$\bm{3.82\pm0.0658}$\\
\cmidrule(lr){3-7}
&&Reconst.
&-&-&$1.73\pm0.167$&$3.94\pm0.131$\\
\midrule
\multirow[c]{4}{*}{\makecell{Flute / Violin / Clarinet\\(Rondeau)}}
&\multirow[c]{4}{*}{128}
&SS--DDSP
&$22.6\pm11.1$&$12.9\pm3.26$&$3.79\pm0.355$&$2.66\pm0.0338$\\
&&DDSPMM
&$16.5\pm2.88$&$7.33\pm0.427$&$2.36\pm0.0792$&$2.87\pm0.154$\\
&&Proposed
&$\bm{16.4\pm3.67}$&$\bm{3.12\pm1.02}$&$\bm{1.98\pm0.202}$&$\bm{3.22\pm0.263}$\\
\cmidrule(lr){3-7}
&&Reconst.
&-&-&$1.74\pm0.171$&$3.41\pm0.216$\\
\bottomrule
\end{tabular}
\label{tab:exp-result}
}
\end{table*}

\section{Experiments}
\subsection{Experimental Setup}
To evaluate the effectiveness of the proposed method, we conducted synthesis parameter estimation experiments on the University of Rochester multimodal music performance (URMP) dataset~\cite{Li2019IEEE_TMM}.
This dataset contains instrument performances on 44 classical pieces (\SI{4.6}{\hour} in total) and musical score information time-aligned with the performances.

We compared three estimation methods: \textbf{SS--DDSP}, \textbf{DDSPMM}, and \textbf{Proposed}.
We also include \textbf{Reconst.} as an oracle reference, obtained by resynthesizing each ground-truth source signal with the pretrained DDSP autoencoder.
For a fair comparison, all methods used the same pretrained DDSP autoencoder ($D=16$) and multi-scale spectral loss settings as in \cite{Kawamura2022ICASSP}.
We divided each mixture into \SI{12}{\second} ($T=375$) segments and applied the estimation methods independently.
 
\textbf{SS--DDSP} applies the pretrained DDSP autoencoder to source signals separated by the score-informed source separation method proposed in~\cite{Montoro2019EURASIP}.
Following~\cite{Kawamura2022ICASSP}, we used it as the separation-and-analysis baseline with the same parameter settings.

\textbf{DDSPMM} performs the DDSPMM-based estimation proposed in~\cite{Kawamura2022ICASSP}.
It serves as a baseline without the DDPM-based generative model.
The $\Fo$ and loudness values were initialized using \Cref{eq:f0_score_initialization,eq:ld_score_initialization}, and the timbre features were initialized by sampling from the standard normal distribution.
For $\lscore_{k,t}$, we set $l_{\text{high}}=-40$ and $l_{\text{low}}=-60$.
The synthesis parameters were updated using the Adam optimizer~\cite{kingma2015adam} with a learning rate of $0.1$ for \num{1000} iterations.

\textbf{Proposed} uses the proposed DDSPMM-guided reverse diffusion algorithm.
We scheduled the guidance weight as 
$\lambda(\tau)=\lambda_{\mathrm{end}}\left((\diffsteps - \tau)/(\diffsteps - 1)\right)^{\omega}$
with $\lambda_{\mathrm{end}}=3.0$ and $\omega=4.0$.
This schedule gradually increases the guidance weight toward the end of reverse diffusion, because applying strong guidance from the beginning often destabilized estimation in preliminary experiments.

Following~\cite{Kawamura2022ICASSP}, we used mean absolute errors (MAEs) of $\Fo$, loudness, and MFCCs extracted from each ground-truth source signal and its synthesized counterpart as evaluation metrics.
We also used the Zimtohrli score~\cite{alakuijala2025zimtohrli} as an objective measure correlated with perceptual similarity between the synthesized and ground-truth source signals.
Higher Zimtohrli scores, ranging from 1 to 5, indicate greater perceptual similarity.

\subsection{DDPM Pretraining}
\label{sec:ddpm-pretraining}
For DDPM training, we used the same isolated instrument recordings used to train the DDSP autoencoder in \cite{Kawamura2022ICASSP}.
From these recordings, we extracted the synthsis parameters using the pretrained DDSP autoencoder and 
split into \SI{12}{\second} segments (652 segments in total).
The selected instruments were violin, flute, viola, cello, clarinet, and oboe ($I=6$).
For the score-informed loudness values, we set $l_{\text{high}}=-40$ and $l_{\text{low}}=-60$, the same as in DDSPMM.

As the noise predictor, we used the U-Net described in \Cref{sec:ddpm-generation} with channel sizes $64$, $128$, and $256$ and a bottleneck width of $512$.
Each convolutional layer had a $3\times 3$ kernel with padding $1$.
The embedding dimension of the diffusion step $\tau$ and the instrument condition $\bm{c}_{\text{inst}}$ were set to 128 and 64, respectively.
The conditioning weight for $\bm{c}_{\text{score}}$ were set to 0.1.
We set $\diffsteps=1000$, matching the number of the fitting iterations in DDSPMM, with the linear noise schedule $\beta_\tau=0.0001+0.0199\times(\tau - 1)/(\diffsteps - 1)$.
We trained the network using the Adam optimizer with a learning rate of $0.001$ and a batch size of $16$ for $2000$ epochs.
The learning rate was decayed by a factor of $0.98$ every $1000$ steps.

\subsection{Results} \label{sec:results}
\Cref{tab:exp-result} shows the means and standard errors of the evaluation metrics over all segments.
Here, $\Fo$ and loudness MAEs for Reconst. are omitted because the corresponding parameters are identical to the ground-truth values.
SS--DDSP yielded larger errors than the other methods in most metrics.
Its $\Fo$ MAE was markedly large for the two-flute mixture, and its standard errors for $\Fo$ MAE were larger than the other methods across all test mixtures.
This trend is consistent with~\cite{Kawamura2022ICASSP}, which showed that separation failures destabilize synthesis parameter estimation and that the analysis-by-synthesis approach provides more stable estimation.

Compared with DDSPMM, Proposed provided substantial improvements in most metrics.
For loudness and MFCCs, it achieved the best performance in all test mixtures.
These metrics reflect synthesis parameters that are not fully specified by musical score information, whereas $\Fo$ can be initialized relatively close to the ground truth from the score.
Thus, the benefit of DDPM-based regularization is more evident for loudness and timbre-related features than for $\Fo$.
The proposed method also gave the highest Zimtohrli scores in most test mixtures, indicating that these parameter improvements led to better perceptual similarity.
Although it did not reach the oracle performance, Proposed reduced the gap between DDSPMM and Reconst. in most cases.
These results demonstrate the effectiveness of DDPM-based regularization for synthesis parameter estimation.

\begin{figure}[t]
\centering
\begin{subfigure}{\linewidth}
    \centering
    \includegraphics[width=\linewidth]{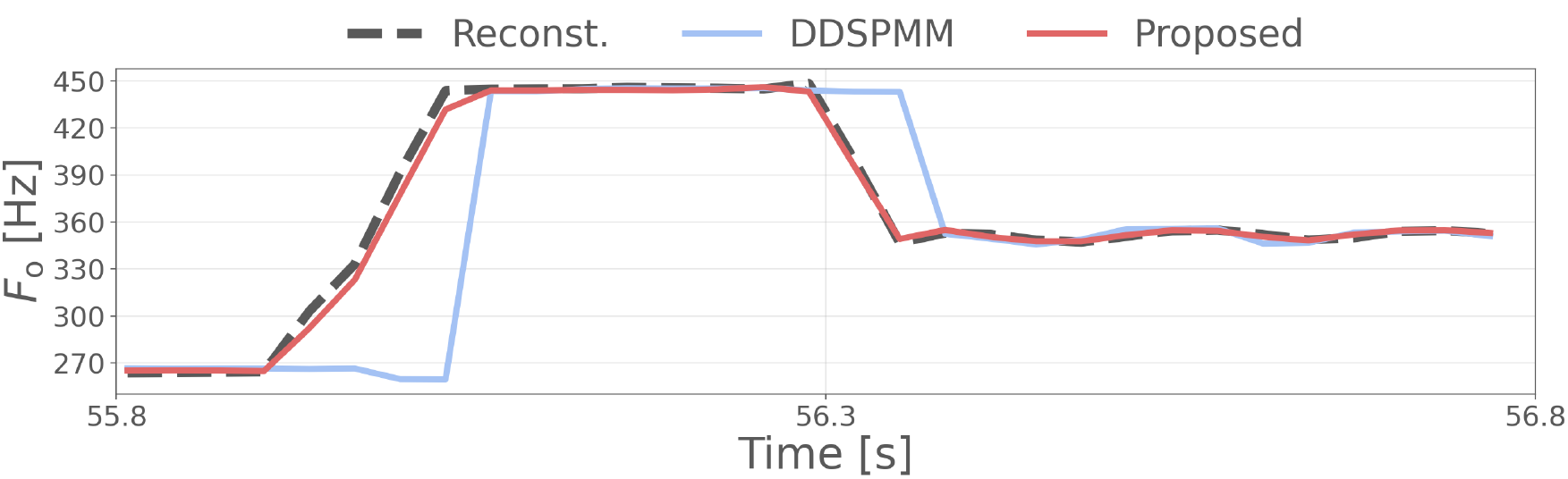}
\end{subfigure}
\\
\vspace{2mm}
\begin{subfigure}{\linewidth}
    \centering
    \includegraphics[width=\linewidth]{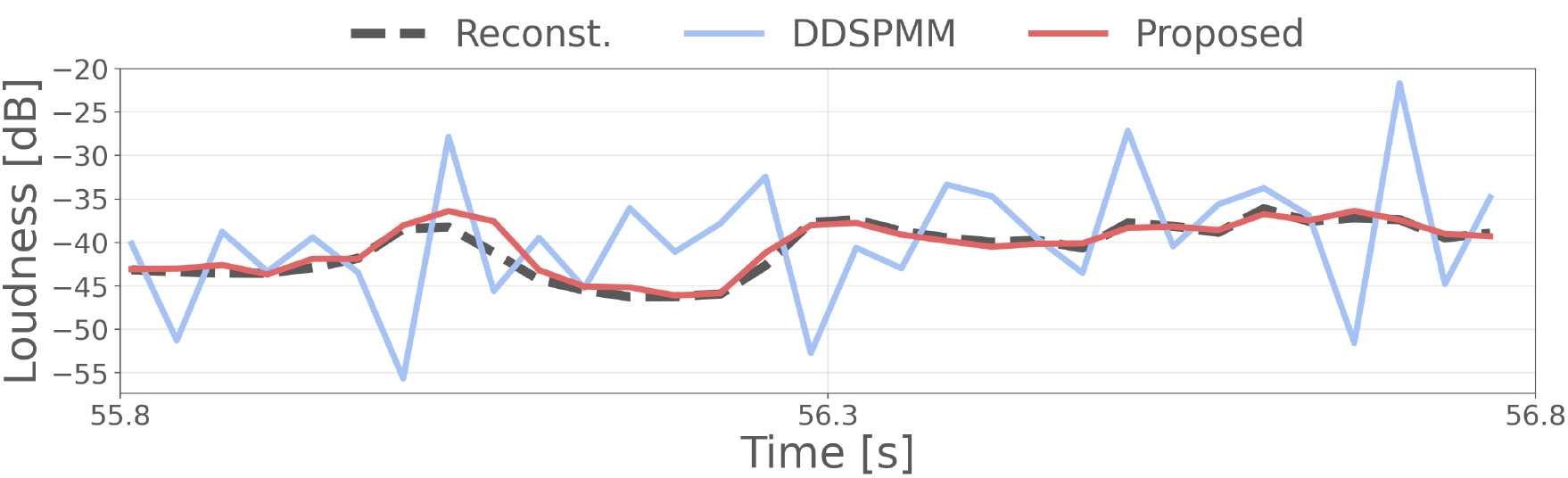}
\end{subfigure}
\\
\vspace{2mm}
\begin{subfigure}{\linewidth}
    \centering
    \includegraphics[width=\linewidth]{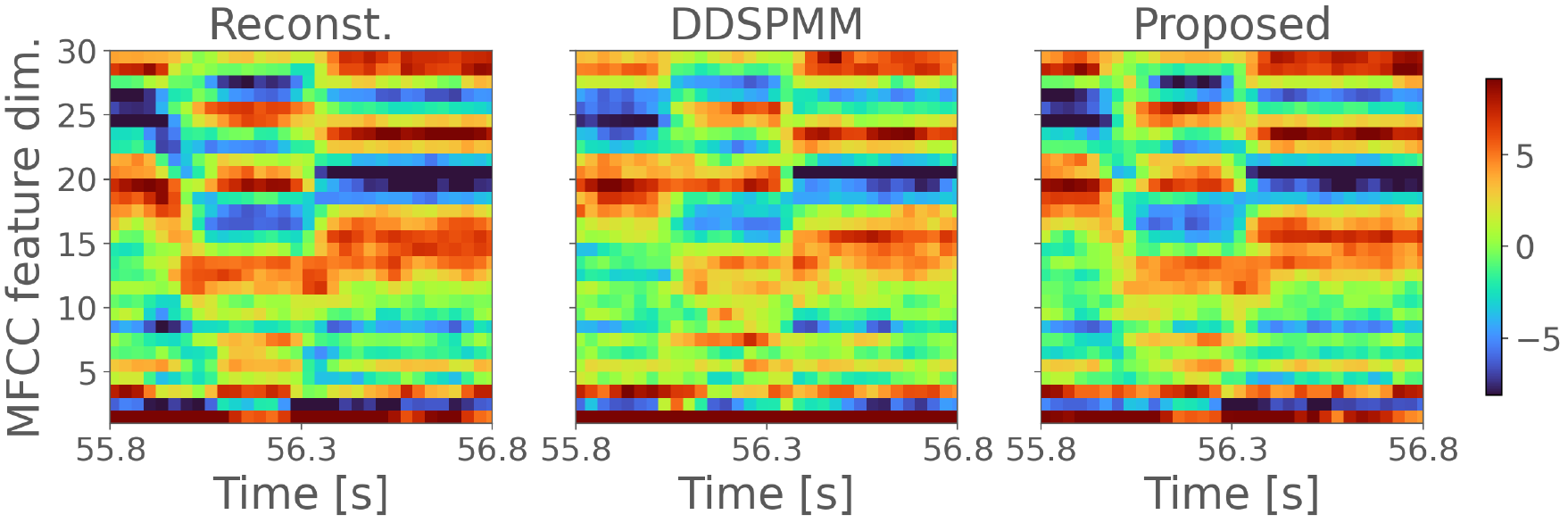}
\end{subfigure}
\caption{Illustrative examples of violin $\Fo$, loudness and MFCCs trajectories estimated from the Flute / Violin / Clarinet mixture in Rondeau. The proposed method suppresses excessive temporal fluctuations while preserving performance-related variations in $\Fo$, loudness, and MFCC.}
\label{fig:synthparam-example}
\end{figure}

\Cref{fig:synthparam-example} shows an example of estimated synthesis parameters.
Compared with DDSPMM, Proposed estimated a smoother and more accurate $\Fo$ trajectory around note transitions.
It also produced a more accurate loudness trajectory by suppressing the excessive temporal fluctuations observed in DDSPMM.
For MFCCs, it yielded temporal patterns closer to the ground truth.
Informal listening suggested that the proposed method reduced unnatural loudness fluctuations and yielded sounds with timbres closer to the ground truth.
These observations show that the proposed method can better estimate temporally structured variations in the synthesis parameters.

\section{Conclusion}
We proposed a method for estimating source-wise synthesis parameters with temporally plausible trajectories from a mixture of harmonic instrument sounds.
The proposed method uses a DDPM as a generative model of synthesis parameters.
During estimation, it guides the reverse diffusion process using the reconstruction error between the observed mixture and the mixture resynthesized by DDSPMM from the current source-wise parameter estimates.
This guidance induces the generated trajectories to be consistent with both the learned synthesis-parameter distribution and the observed mixture.
Experiments showed that the proposed method improved estimation accuracy, especially for loudness and timbre-related features.

\printbibliography

\end{document}